# Analysis of the Visually Detectable Wear Progress on Ball Screws

Tobias Schlagenhauf, Tim Scheurenbrand, Dennis Hofmann, Oleg Krasnikow

*Abstract* — The actual progression of pitting on ball screw drive spindles is not well known since previous studies have only relied on the investigation of indirect wear effects (e. g. temperature, motor current, structure-borne noise). Using images from a camera system for ball screw drives, this paper elaborates on the visual analysis of pitting itself. Due to its direct, condition-based assessment of the wear state, an image-based approach offers several advantages, such as: Good interpretability, low influence of environmental conditions, and high spatial resolution. The study presented in this paper is based on a dataset containing the entire wear progression from original condition to component failure of ten ball screw drive spindles. The dataset is being analyzed regarding the following parameters: Axial length, tangential length, and surface area of each pit, the total number of pits, and the time of initial visual appearance of each pit. The results provide evidence that wear development can be quantified based on visual wear characteristics. In addition, using the dedicated camera system, the actual course of the growth curve of individual pits can be captured during machine operation. Using the findings of the analysis, the authors propose a formula for standards-based wear quantification based on geometric wear characteristics.

## I. Introduction

Ball screw drives (BSD) are among the most commonly used components for the conversion of rotary into translational motion in modern machine tools (Forstmann, 2010). The operating time to failure of BSDs varies significantly in different environments depending e.g., on the respective load spectrum (Fleischer et al., 2013). Without a reliable wear quantification system, this can result in unforeseen component failures, leading to unplanned downtime, increased repair times, and, in particular, to higher costs (Dehli, 2020). Therefore, maintenance has subsequently changed from former reactive to modern preventive, predictive, and even proactive approaches (Stamboliska et al., 2015). Consequently, condition monitoring technologies are complemented by advanced statistical methods for the estimation of reliable component lifetimes and time-of-failure predictions. This presupposes a reliable detection and quantification of the current wear state and progress. In the context of BSDs, there are several approaches to this, all of which are based on the detection of indirect effects caused by advanced wear and tear. The objective of this paper is to present the first direct time-continuous measurement approach for quantification of the current wear state on BSD spindles. The images resulting from the here proposed approach might lead to less measurement uncertainty, high spatial resolution, and good visual interpretability of the results by maintenance staff. Therefore, an analysis of the visually detectable wear progress on BSD screws based on an integrated camera system developed by (Schlagenhauf et al., 2019) and (Schlagenhauf et al., 2020) is carried out. For this purpose, ten BSD spindles are systematically worn out under controlled conditions in a suitable test facility at the wbk Institute of Production Science at KIT, Karlsruhe. The wear progress starting with the early emergence of pits and proceeding to the ongoing spread of pitting is detected and analyzed until the final failure of the component. For the first time, the growth of defects on BSDs is mathematically described based on images. From this, a standards-based visual wear quantification approach is proposed and validated.

## II. Related Work

### A. Wear Phenomena on Ball Screw Drives

Initially (Haberkern, 1998) defines the three mechanisms of *sudden early failure*, *slow loss of preload,* and *late failure* as the underlying principles of BSD failure. *Sudden early failures* are caused by plastic deformations in the BSD-System (Haberkern, 1998), (Spohrer, 2019). Because of the sudden occurrence of sudden early failures, a visual, time-continuous observation is not possible. *Slow loss of preload* is caused by an abrasion effect of the BSD balls as a result of a constant lapping process due to the smallest vibrations or constant short-stroke operation (Haberkern, 1998). Consequently, this is not a wear phenomenon of the BSD spindle itself and will therefore not be considered in detail in this paper.

According to (Haberkern, 1998) and (Forstmann, 2010), the third mechanism, *late failure*, is caused by periodic loads, as they frequently occur in rolling or pitching contacts. This wear mechanism leads to the wear phenomenon known as pitting. Even though (Haberkern, 1998) states that slow late failure hardly occurs in industrial scenarios with gentle operating conditions, we show that it happenes under high but realistic loads. According to (Haberkern, 1998) and (Imiela, 2006), tribological wear mechanisms even occur on a properly assembled and lubricated BSD since the lubrication film is regularly interrupted due to reversing operating modes. (Sommer et al., 2018) state that many wear processes involve a sequence of different wear mechanisms since the basic wear mechanisms adhesion, abrasion, surface disruption, tribochemical reaction, and ablation rarely occur individually from each other. These observations are also supported by (Helwig, 2018) who investigated the wear characteristics on ball screws and expressed the assumption of mutually influencing wear mechanisms. (Sommer et al., 2018) explain this by detached wear particles, which can be a starting point for further wear when a ball rolls over them. Accordingly, (Schopp, 2009) conducted structure-borne noise analyses on ball screws during destruction tests and derived a three-phase wear curve of BSDs depicted in Figure 1 which was also presented by (Munzinger et al., 2009) and (Spohrer, 2019).

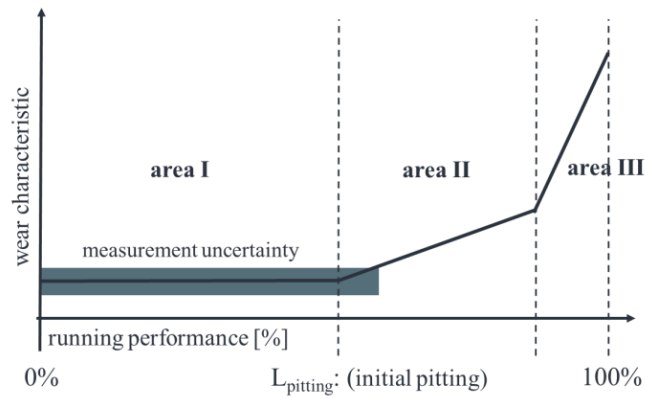

*Figure 1 Wear curve (structure-borne noise) for BSD (schematic) according to (Schopp, 2009).*

Initially, the wear signal (e.g., structure-borne noise) is constant, since no to minimal wear occurs (area I). With increasing running performance, the measured signal increases linearly, indicating the formation and progression of wear (area II). (Schopp, 2009) concludes that wear must be present in the system from approximately 60 % (= $L_{pitting}$) of the service life. The gradient of the curve rises sharply at approximately 90 % of the service life due to mutually reinforcing wear mechanisms (Forstmann, 2010) and finally culminates in the mechanical failure of the system (area III). (Schopp et al., 2009) emphasize that a reliable wear prediction from structure-borne noise is inaccurate before reaching 60 % of the service life. Furthermore, the noise of the sensor signals and thus the uncertainty increases with advancing service life. It can thus be stated that, although a damage condition has basically been captured by structure-born noise, the exact wear progression has not yet been represented in image data.

### B. *Service Life Standard for BSD*

The service life of BSD is defined by (DIN ISO 3408-1, 2011, p. 13) as "the number of revolutions that a ball screw nut (or a ball screw) performs with respect to the ball screw (the ball screw nut) before the first signs of material fatigue appear on either the screw, nut, or rolling element" [transl. by the authors]. An empirically derived calculation rule for the nominal life $L_{10}$ is defined by (DIN ISO 3408-1, 2011) as well. The rule dates back to (Lundberg & Palmgren, 1952) and is based on the dynamic axial load rating $C_a$. Based on the service life, the failure of a component is defined according to one exact point in time (the point in time when wear occurs), while the $L_{10}$ is more of a statistical statement about the minimum time, that 90% of ball screws can be operated under the same load. The nominal lifetime is calculated with: $L_{10} = (\frac{C_a}{F_m})^3 * 10^6 \ [Revolutions]$. The service life calculation also depends on the used type of nut. Criticizing the service life definition, (Huf, 2012) and (Klein & Brecher, 2011) state that reaching the service life does not necessarily result in the practical inoperability of the system. Additionally, (Schaeffler, 2000) describes that after the occurrence of a defect, the system can usually still be operated for a certain amount of time. (Drescher, 2015), (Imiela, 2006), and (Münzing, 2017), on the other hand, describe that ball screws often have to be replaced before reaching their theoretical service life due to contamination in the system, incorrect operation, or excessive wear. In addition, (Fleischer et al., 2013), (Haberkern, 1998), and (Schopp, 2009) observed service lives of ball screws that were significantly above, as well as below, the calculated nominal service life. (Haberkern, 1998), (Münzing, 2017) and (Spohrer, 2019) classify the calculation rules for the $L_{10}$ life of ball screws given in (DIN ISO 3408-5, 2011) as being insufficient for real applications. (DIN 631, 2020) defines the service life of profile rail guides, which have a lot in common with BSDs. This standard has already been used by (Münzing, 2017) as the foundation for a novel definition of service life for BSD. According to this, the end of service life is reached as soon as a pitting zone on the raceway exceeds 0.3 times the ball diameter. As a function of the rolling element diameter $D_w$ and the diameter of a damage (pitting) $d_s$, it holds: $d_s \geq D_w * 0.3$. However, to use this definition, continuous and quantitative monitoring of the damaged areas on the entire BSD spindle surface is necessary. A fact not taken into account by this formula is that there is no specific service life limit, which is generally valid for all applications. Instead, depending on the respective accuracy and smoothness requirements, a BSD can often be used beyond the occurrence of pitting of a certain size.

### C. *State-of-the-art Wear Detection Approaches for BSD*

Wear analysis approaches for BSDs are divided into load-based and condition-based approaches (Munzinger et al., 2010), (Hennrich, 2013). Load-based approaches estimate the wear condition of a component based on the cumulated loads applied to it during operation (Hennrich, 2013). They can be measured effortlessly during operation but lead to a reduced prediction quality (Broos, 2012), (Huf, 2012). Condition-based approaches, on the other hand, determine the current wear condition based on actual wear characteristics (Hennrich, 2013). They are either measured directly or indirectly based on secondary variables. The state-of-the-art wear quantification approaches for BSDs mainly utilize indirect measurement principles based on the detection and quantification of wear effects, rather than the wear characteristics themselves. (Hennrich, 2013), (Helwig & Schütze, 2018), (Xi et al., 2020), (Schmid et al., 2010), and (Schopp, 2009) present structure-borne noise-based approaches. (Veith et al., 2020), (Herder, 2013), (Möhring & Bertram, 2012), and (Imiela, 2006) utilize the pre-tension loss in the nut caused by wear. The loss of pre tension is a relevant quality measurement and could also be regarded as relevant as the late failure which is considered in this work. Though the here presented approach has advantages in terms of interpretability of the specific damage feature pitting. (Riaz et al., 2021), (Riaz et al., 2020), (Spohrer, 2019), (Yagmur, 2014), (Hennrich, 2013), (Verl et al., 2009), (Q. Yang et al., 2020) and (Cipollini et al., 2019) make use of the control internal motor current to indicate wear in the system. (Münzing & Binz, 2017), (Broos, 2012), (Stockinger, 2011), and (Imiela, 2006) presented model-based approaches, and (A.

Mannesmann, 2021) measures the temperature on the flange of the nut to quantify wear. (Cheng et al., 2019) use the temperature in the roller bearings of train wheels for the monitoring of their condition. Additional approaches, using the temperature of gear units can be found in (Touret et al., 2018).

(Okwudire & Altintas, 2009) use a finite element model incorporating the axial, torsional and lateral dynamics of the Ball Screw to increase the modelling of the positioning accuracy. This approach enables engineers to better plan the fatigue life of an BSD in high speed applications. (Jia et al., 2019) use the motor current of a feed axis to extract the statistical ling term characteristics of the system, which are then used to improve the health assessment of the system. (Xi et al., 2020) present a model to monitor the loss of stiffness of a ball screw drive based on the motor current. (Benker et al., 2019) use vibration signals of the BSD to build a probabilistic classification model to make decision regarding wear in the BSD-System. (H. Yang et al., 2021) predict the remaining useful lifetime of BSDs by integrating the backlash of the BSD into a stochastic degradation model. (Hinrichs et al., 2021) analyze the impact of compressing data for the use in algorithms for the condition monitoring of BSD. (Alqatawneh et al., 2021) successfully investigate the use of neural networks for the condition monitoring of a transmission based on structure-borne noise. (Krishnakumar et al., 2018) investigate several methods based on machine learning techniques to analyze vibration data of a machine tool for prognostics health management of cutting tools. They show that the use of a neural network is best suited for this task. (Lee et al., 2021) study the application of deep learning techniques on vibration data of a roller bearing. As a result they could match the wear state of the bearing with an accuracy of > 90%. (Serin et al., 2020) give a detailed review about the use of deep learning techniques for the analysis of motor current, structure-borne noise and airborne sound for condition monitoring purposes. Further work on challenges regarding the analysis and control of machine tool feed drives is presented by (Altintas et al., 2011) which review the characteristics of several common machine tool feed drive systems. (Butler et al., 2022) give a recent overview of condition monitoring approaches for machine tool feed drives.

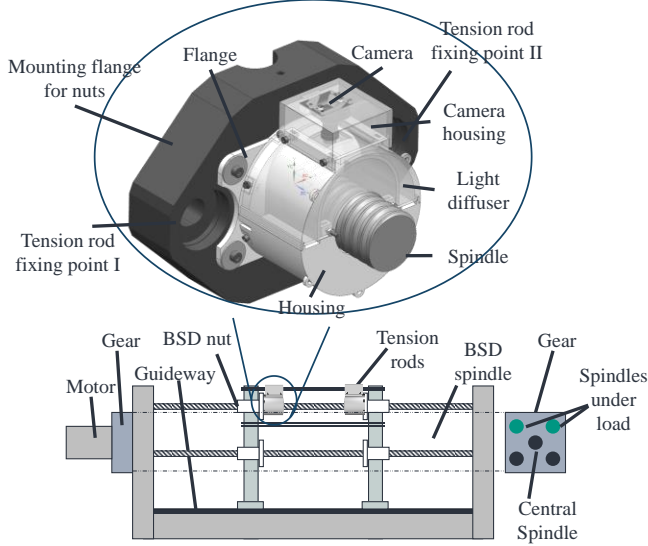

*Figure 2: Setup of the test bench and detailed view of the integrated camera system. Best viewed on screen.*

Given the current state of research, there is no well-known direct measurement approach and, in particular, none that is image-based.

### D. *Camera System for Wear Analysis on BSD*

The foundation for the here presented visual wear progression analysis is an image dataset representing the whole lifecycle of multiple BSDs. The dataset contains images of the spindle surfaces in a high time and spatial resolution. It is generated using a camera system presented in (Schlagenhauf et al., 2019). The system as depicted in Figure 2 mainly consists of an *OV5647* camera, a light source with a diffuser, and a communication interface. The assembly can be mounted to various ball screw nuts using interchangeable mounting adapters. The enclosed housing prevents the penetration of contaminants and chips as well as ambient light influences from the outside. As shown in (Schlagenhauf et al., 2019), this guarantees a consistent image quality. A dataset is generated in an iterative process by repeatedly performed camera drives. During a camera drive, the BSD spindle is rotated until the camera unit has traveled the area of interest of the spindle. Samples of an image dataset generated in this way are shown in . The three images are taken from the same spindle spot at consecutive points in time. Preliminary tests have shown that the image quality remains mainly constant throughout the entire test and that contamination, e.g., by the lubricant, has only a small effect on the image data.

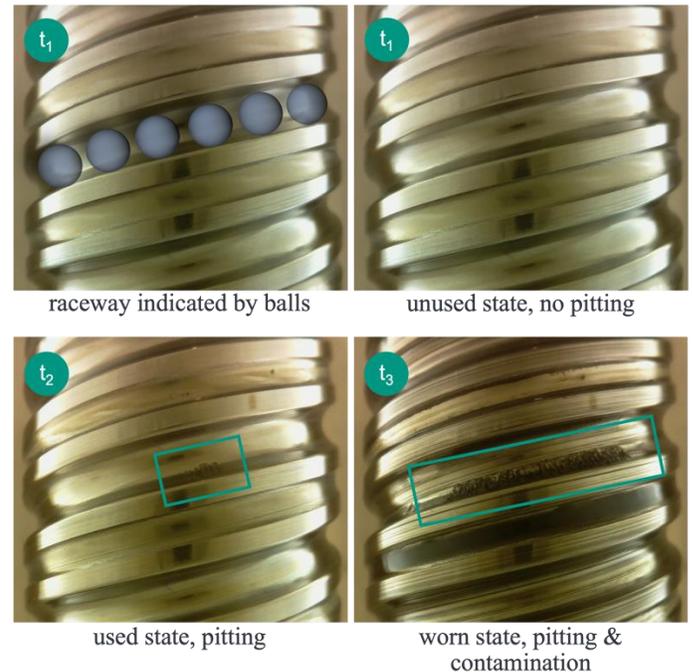

*Figure 3: Three consecutive spindle states $t_1$ - $t_3$ from the image dataset generated by a camera system for BSD*

## III. OWN APPROACH

According to the state of the art, there is a variety of indirect measurement variables and methods proven to be suitable for wear detection. As shown in various papers (see II.C), different wear mechanisms and effects can be detected particularly well by distinct measurement methods. In contrast to these indirect approaches, image data offers the opportunity for direct, visual wear detection. The objective of this work is to understand the evolution of pitting on the surface of BSD spindles. Therefore, the visually detectable wear phenomenon *pitting* is evaluated regarding defined wear parameters to quantify the wear progression on BSDs. In addition, a formula based on existing component lifetime standards is defined as a suggested termination criterion for condition-based maintenance of BSDs. To this end, the wear progression from the initial emergence of pitting, through the superposition of wear mechanisms, to the final failure of a BSD is examined using images of spindle surfaces. The image dataset is generated in destruction tests at the Institute of Production Science at the Karlsruhe Institute of Technology.

### A. *Experimental Setup*

The destruction tests are performed on a test bench which can wear up to five spindles at a time. In the presented tests, a total of ten spindles are worn under constant conditions in five runs of two spindles each. The spindles are Bosch Rexroth Spindles with a diameter of 32 mm, a lead of 20 mm, and a ball diameter of 3.969 mm. The product name which can be used in the configurator (Rexroth, 2022) is *BASA / 32x20Rx3.969 / FEM-E-B - 3 / 02 / 1 / 2 / T7 / R / 21KXXX / 21KXXX*. The Ball Screw drives have a dynamic axial load factor $C_a$ of 23.6 kN with a correction factor of 0.9 and a pre-tension-class of C3. Two BSD nuts are mounted on each spindle, together with the camera system presented in chapter II.D. The two nuts are connected by tension rods to apply a specific axial tension force. The axial forces are set considerably high ($0.4*C_a$ at 20°C which results in ~ $0.6*C_a$ at an operating temperature of 50°C) to obtain realistic results in a reasonable test time. The experiments are run with a constant speed of 400 revolutions per minute. The wear tests terminate as soon as a critical temperature of 70 C is exceeded at a BSD nut flange. The value of 70°C is based on (skf, 2022) where it is stated that the allowed operating temperature of classical BSDs is up 110°C. Since the authors measure the temperature at the nut flange, an maximum allowed temperature of 70°C is chosen. The BSDs are lubricated following manufacturer's specifications. The experimental setup is depicted in Figure 2. A camera drive is automatically triggered every four hours during the entire wear test (= time resolution). Within a camera drive, the images are recorded with the spindle being rotated by 22.5 ° in between each one (= spatial resolution), thus capturing the entire spindle. Two spindles with two nuts each are mounted. Through a tension rod, which connects the nuts, the nuts are loaded with the load described above ($0.4*C_a$). The central spindle is load free and connected with the motor. The loaded spindles follow the central spindle via a gearbox (gear in Figure 2). The tension rods are not directly acting on the nuts but on the adapters on which the nuts are attached. The flanges for the nuts are constructed in such a way, that no tilting moments can occur at the nuts. Additionally, the loads at the tension rods are held at the same level. This ensures an even load distribution. The size of the pittings was extracted manually from the image data. By knowing the spindle diameter and the resolution of the images, a conversion to mm$^2$/pixel was made. The analysis of the area was then done using the software GIMP.

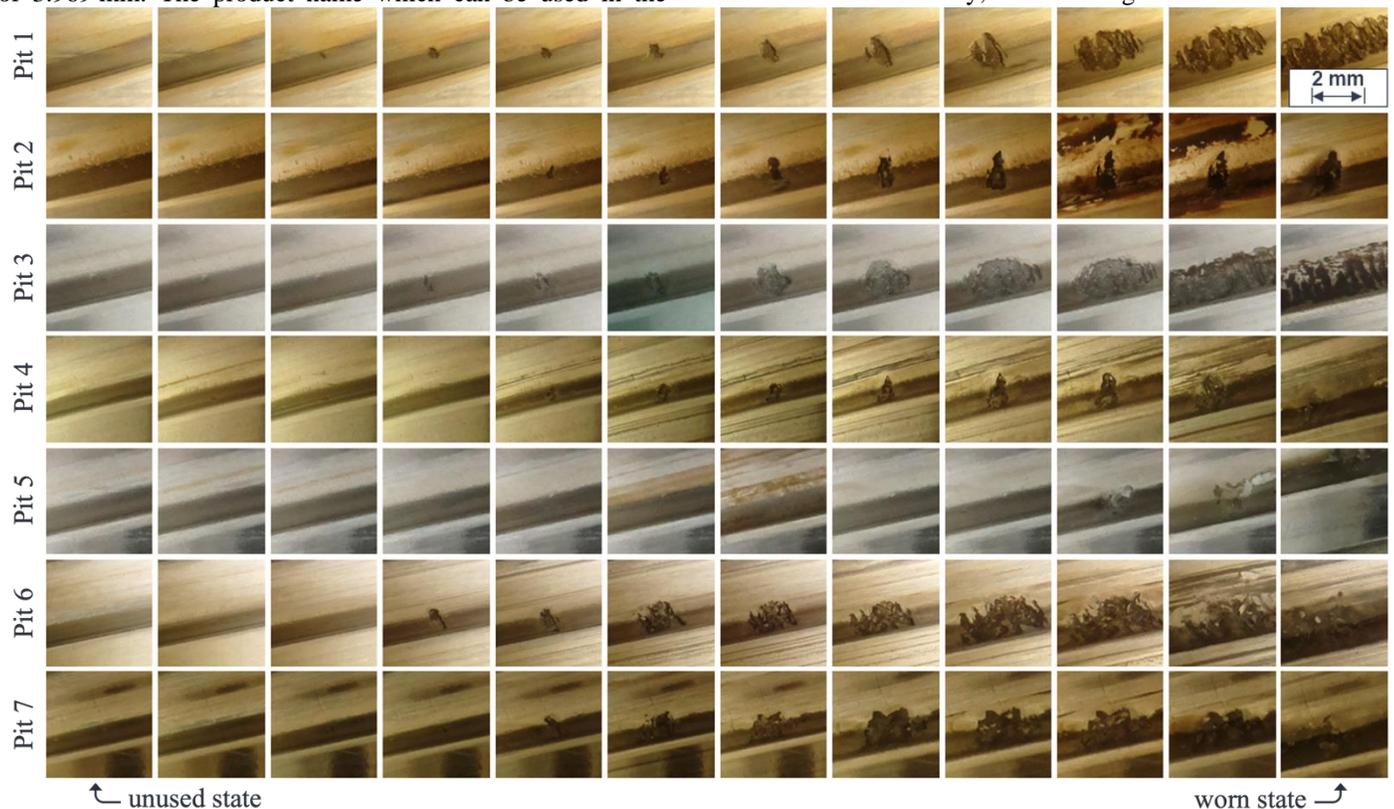

*Figure 4: Exemplary growth of seven randomly picked pits from multiple BSDs. The first column shows the initial state, the last column shows the worn state.*

### B. *Pitting Image Dataset*

The pitting image dataset is generated from a total of 500.989 images taken during the destruction tests. During the experiments, 148 pittings were located on the surfaces of the spindles. The growth of each pitting is depicted in the image data. As a result, the pitting image dataset represents the entire damage history of each location on the spindles where pitting occurred. Each 24-bit sRGB image has a resolution of 2592 x 1944 pixels.

## IV. RESULTS AND DISCUSSION

The analysis of visually detectable wear is performed using the image dataset presented in the previous chapter III.B. Figure 4 shows seven randomly selected wear progressions from the dataset. The first column depicts the installation condition of the spindle, while the last column shows the worn surface sections at the end of the wear tests. The effect of heavy contamination and debris on the visibility of pitting is found to be largely negligible in the aggregated data. This is evident in Figure 4 as well. The images are evaluated with respect to the following parameters: Axial length, tangential length, and area of pitting for each timestamp and pitting location. In addition, the total number of pits for each timestamp and the time of initial visual appearance of each pit are considered. The results of the empirical analysis will be presented in the following.
During the destruction tests, the formation and growth of a total of 148 pitting areas are observed on the surface of eight spindles. Two out of ten spindles show no visually detectable wear (BSD 9 & 10) although they were strained with the same loads as BSD 7 & 8, which show a multitude of pits. The results show no clear evidence why there are not defects on those two spindles. Due to the termination criteria of 70°C, it was not possible to investigate whether pitting would have occurred at a later stage. Though this could be assumed as likely. As depicted in Figure 5, the spread in the amount of pitting is also reflected in the achieved lifetimes of the analyzed BSDs. The box plot shows the observed lifetimes of the ten spindles normalized to their nominal lifetimes. The median of the spindle's service life is on average significantly higher than the nominal lifetimes according to (DIN ISO 3408-5, 2011). At the same time, however, an extreme spread of the lifetimes can be observed: One spindle has already failed at about 50% of the $L_{10}$, while other spindles have reached more than ten times $L_{10}$ without any visible pitting.

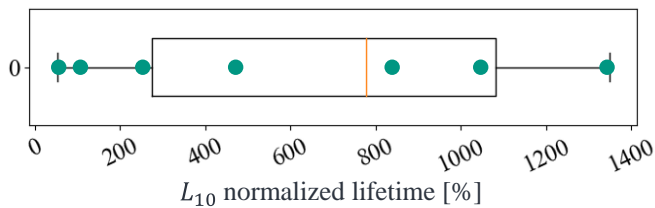

*Figure 5: Locality, spread, and skewness groups of the observed spindle lifetimes normalized to their calculated nominal $L_{10}$ lifetimes.*

The strong discrepancies in both the number of pits and the observed lifetimes underline the additional value that reliable, continuous wear monitoring provides compared to a classical lifetime calculation alone.

### A. *Amount of Pitting Spots*

Figure 6 depicts the number of pits for each BSD over the normalized lifetime of each spindle.
Since the absolute lifetimes of the spindles vary significantly

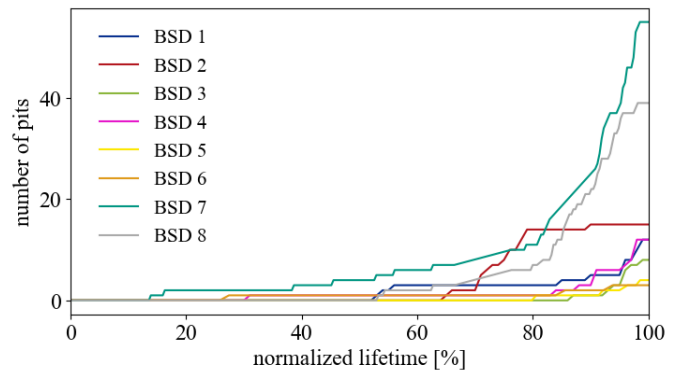

*Figure 6: Time progression of the amount of pitting spots on the surface of the spindles.*

(Figure 5) and the main interest of the authors lies on the relative growth of the pittings, the course of each pitting was normalized. Here, 100% represents a worn spindle. From now on, the normalized lifetime is used to describe the lifetime of a pitting on the spindle with respect to the component failure. Following (Schopp, 2009), the wear development of the BSDs can be divided into three phases. After an initial period without pitting, the first phase of minor initial pitting starts at approx. 20 % of the BSD's lifetime. Characterized by an increased emergence of pitting, the second phase starts at approx. 60 % of the lifetime. The highest rate of pitting is observed in the third phase starting at approx. 80 %, therefore resulting in a strong increase in pits. (Forstmann, 2010) assumes that wear particles influence each other, from which he develops the theory of mutually reinforcing wear mechanisms (e. g. wear particle-enriched lubricant acts like an abrasive medium). For the first time, this assumption is visually supported by the analysis of the image data, as all eight BSDs affected by pitting experience a high increase in defect incidence towards the end of their lives.

The number of pitting areas alone only provides a limited indication of the BSD's wear progression. Therefore, following the growth of the pitting areas is examined in more detail.

### B. *Surface Area of Pitting*

Figure 7 depicts the total surface area of pitting on each spindle. Again, a three-phase progression is observed for 6 of the spindles, where this is especially obvious for four of the spindles. The total pitting area barely increases up to approx. 60 % of the lifetime (area 1). From 60 % to 80 % a slow increase in the total pitting area is observed (area 2). Exceeding approx. 80 % of the lifetime, a sharp increase in the total pitting area is observed (area 3). Though, since this is not totally clear for all of the spindles, further experiments are necessary to strengthen these results.

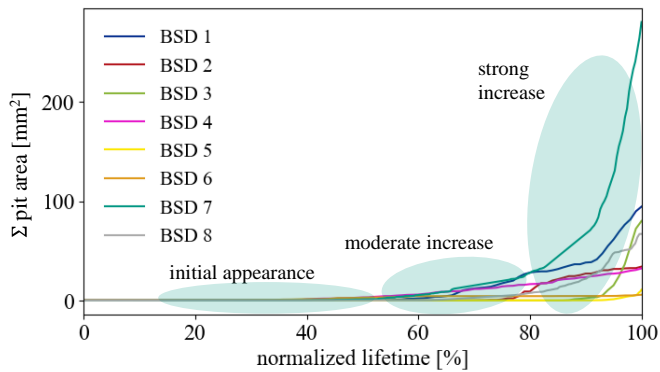

*Figure 7: Time progression of the total area of pitting on the surface of the spindles*

Two phenomena arising mainly at the last 20 % of a BSD's lifetime are hypothesized: First, the increased emergence rate of pits in the third phase leads to many new pits (see area 3, Figure 6). Second, these late-arising pits have a significantly higher growth speed due to mutually reinforcing wear mechanisms as stated by (Forstmann, 2010). This is also coincident with Figure 8, which depicts the temporal development of each pit's surface area. Each of the 148 curves illustrates the surface area evolution of a pit. For the sake of readability, the legend has been omitted from the plot. The exact course of the pittings is not relevant at this point. What is rather relevant, is the fact that the majority of the pits emerge after passing 80 % of the lifetime, and the growth speed increases significantly towards the end of the BSD's lifetimes. This underlines the hypothesis of mutually influencing wear mechanisms.

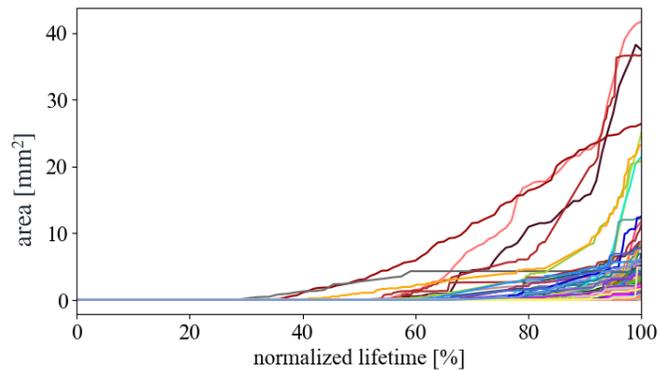

*Figure 8: Time progression of the surface area of 148 pits*

### C. *Geometric Dimensions*

Pitting only occur on the side of the ball raceway that is subjected to strong surface pressure by the passing balls as a result of the axial load applied to the BSD nuts during the tests. Since pitting is directly related to ball contact, it is reasonable to assume that the characteristics of pitting growth differ spatially. Therefore, the upcoming chapter analyzes the pitting growth in two directions: Parallel to the raceway direction (= tangential spindle direction) and orthogonal to the raceway direction (= axial spindle direction).

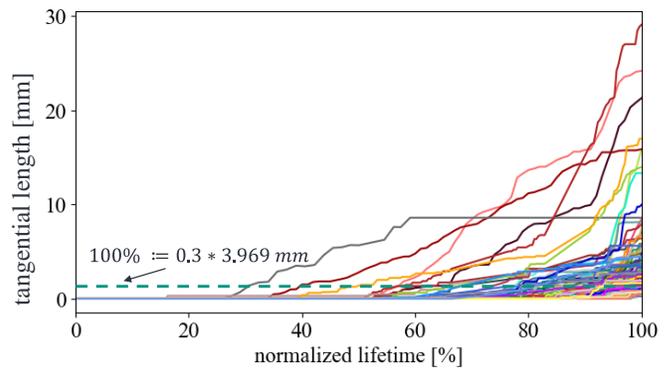

*Figure 9: Time progression of the tangential length of pits. The 100% mark corresponds to the formulation of the service life limit in chapter F. Outlook: Standards-based Wear Quantification.*

Figure 9 depicts the pitting growth in the tangential direction. Once again, it is observed that the growth rate rises with increasing lifetime. Although the qualitative trend is similar to the pit's surface area (Figure 7), the gradient of the curves does not increase to the same extent as previously observed. The primary reason for this is the strongly increasing appearance rate of pits towards the end of the BSD's lifetime, which leads to a disproportionate rise in surface area. The figure also depicts the mark for the life time of the component discussed in section F. It can be seen that the components can be operated beyond reaching the defined 100% lifetime mark. A crucial observation is the absence of a fixed growth limit in Figure 9. This becomes particularly clear when considering the axial expansion. According to Figure 10, the growth of pits orthogonal to the raceway, on the other hand, starts with a phase of rapid growth, followed by a phase of flattening, linear growth. In contrast to the tangential expansion of the pitting surfaces, the axial expansion is limited by the geometry of the balls of the BSD. Therefore, at the end of the wear test, the maximum axial expansion of the pitting areas reaches just one-tenth of the maximum tangential expansion. This effect is also depicted in Figure 10.

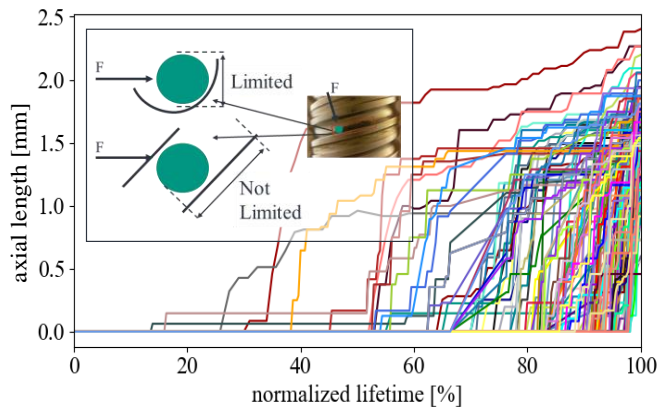

*Figure 10: Time progression of the axial length of pits. Best viewed on screen.*

### D. *Mathematical Representation of Pitting*

A key requirement for a visual-based approach to wear quantification on BSDs is a mathematical description of pits based on geometric features. To this end, we suggest an elliptic approximation of a pit's surface area due to its axial boundary.

The surface of a pit is therefore given by: $A_{pit} = \frac{1}{2}a \cdot \frac{1}{2}b \cdot \pi$ with $a$ as the axial length and $b$ as the tangential length of the pitting. At a very early stage of growth, the length of the major and minor axes can be assumed to be equal ($a = b$), which is also evident from the top row of Figure 11. The figure shows a randomly picked course of pit growth from the image dataset. With this assumption, the circular surface area is given by $A_{pit,init} = \frac{1}{4}b^2 \cdot \pi$ with $b = length\ in\ raceway\ direction$, which is a special case of an ellipse.

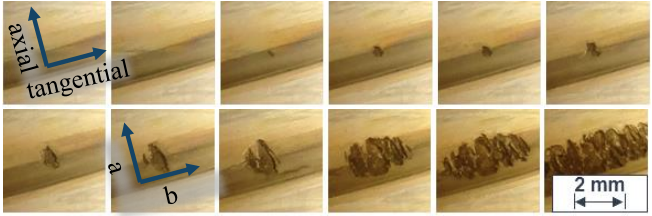

*Figure 11: Typical chronological progression of a pit*

This very early stage of growth is followed by a short phase with the axial length of a pit exceeding the tangential length due to the higher initial growth speed in the axial direction (see Figure 10). However, towards the end of the lifetime the pit's axial expansion is bounded by the fixed ball diameter, described by the $ball-dependent\ constant\ c$: $A_{pit} = \frac{1}{2}b \cdot c \cdot \pi$. Accordingly, the pit's surface area $A_{pit}$ is proportional to its $tangential\ length\ b$. An empirical evaluation of the proposed elliptic approximation was successfully performed. Figure 12 illustrates the results of the evaluation vicariously for four randomly picked pits. The figure shows the area as well as its elliptical approximation over the normalized lifetime. For the calculation of the ellipse area, the tangential (b) and axial (a) extent of the pittings were measured from the images. The area was calculated using the formula $A_{pit} = \frac{1}{2}a \cdot \frac{1}{2}b \cdot \pi$. As can be seen in Figure 12, the approximation matches closely with the empirical results.

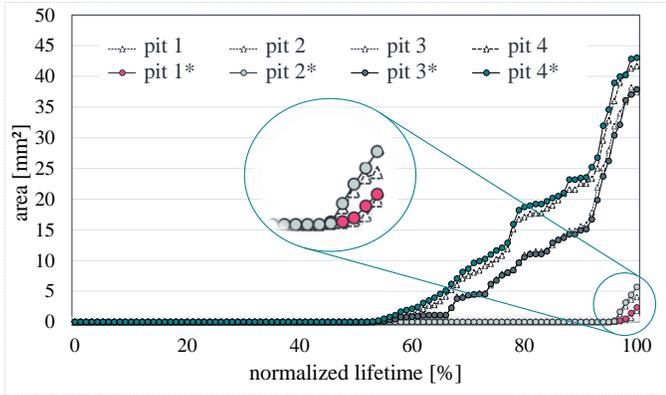

*Figure 12 Comparison of the empirically determined area of four random pitting areas (n) with their elliptic approximations (n\*).*

### E. Comparison of Wear Identification Approaches

State-of-the-art wear quantification approaches for BSDs are usually based on indirect measurements of wear effects, such as pre-tension loss, structure-borne noise, or temperature measurements (see II.C). The camera-based approach applied in this paper allows for direct visual analysis of explicit wear features such as geometric characteristics of pitting in the image data. For the first time, this allows to quantitatively confirm the assumptions about the damage progression, which was previously derived qualitatively from indirect measurement methods. However, in comparison to indirect measurement methods using image data, wear can reliably be detected at a very early stage of development. For instance, structure-borne noise-based approaches as in (Schopp, 2009), (Schmid et al., 2010), (Hennrich, 2013), (Helwig, 2018), or (Xi et al., 2020) are only able to reliably detect wear when approximately 60 % of the component's lifetime is exceeded. Indirect approaches primarily quantify the overall condition of the spindle as a superposition of all wear spots. However, using images of the spindle, with the here presented approach it is possible to quantify the growth of each individual pit over its entire progression. This allows the wear condition of the spindle to be quantified with high accuracy at a high spatial resolution. Additionally, in contrast to indirect approaches (e.g., structure-borne noise) the image data allows an intuitive interpretation of the wear condition by domain experts.

### F. Outlook: Standards-based Wear Quantification

The previous considerations clearly show that the point in time of the first pitting occurrence $L_{Pitting}$ (Figure 6), as well as the normalized lifetime $L_{10}$ (Figure 5) can vary significantly between different BSDs, even though they were operated under the same conditions. At the same time, both $L_{Pitting}$ and the dimensions of the pitting areas are well-suited parameters for quantification of the late failure. At hits point, it has to be emphasized that the presented approach does only consider pits occurring on the BSD spindle, since the camera system is not able to consider the BSD-nut or the BSD-rolling-elements. As demonstrated in the previous sections, $L_{Pitting}$ and the dimensions of pits can be precisely and reliably determined using an image-based visual approach. In the following, a formal approach utilizing this data for reliable condition-oriented wear quantification on BSDs is presented. The approach is inspired by the service life standard for profile rail guides (DIN 631, 2020). This standard was already used by (Münzing, 2017) in the context of BSDs and defines the end of the service life of profiled rail guides as the pitting area diameter $d_s$ exceeding 0.3 times the ball diameter $D_w$: $d_s \geq D_w * 0.3$. Here, $d_s$ denotes the major axis of the elliptical pitting surface, i.e., the tangential pitting length (see Figure 11) that is parallel to the ball raceway. Since one BSD is likely to have many pitting zones, the pitting ellipse with the longest major axis is decisive for the entire spindle. This is a simplified assumption made here. In future work, the sum of all expansions could be considered instead of to the largest expansion. However, considering the largest expansion has proven to be sufficient in this case. To ensure that the component is replaced as late as possible within the scope of its application requirements, while at the same time complying with company-specific safety factors against unexpected spontaneous failure and time lead for maintenance work, it is proposed to add an individual scaling factor α to the definition:

$d_s \geq D_w * 0.3 * \alpha$ with $d_s$: max. pitting Ø, $\alpha \in \mathbb{R}^{>0}$: individual scaling factor, $D_w$: ball Ø. The parameter α can be used to scale the allowed size of the largest pitting. If it is wanted that a component should be replaced as soon as there is a pit, α can be set to small values e.g. 0.1. Setting the value of α to 1 results in the formulation given by (DIN 631, 2020) while setting α to values larger than 1 allows a longer operation of the spindle. It has to be emphasized, that the value of α has to be chosen company and application dependent. Large values for α allow longer operation times but also increases the probability of a breakdown of the component. Smaller values for α results in the opposite. As a result, the service life limit is not absolute but depends on the component properties ($D_w$) and individual characteristics (α), thus leading to a more adaptable wear quantification. As shown in the experiments, BSDs can in fact be used to a limited extent beyond the occurrence of initial pitting. This fact is reflected by the proposed lifetime estimation. However, it should be emphasized that this service life definition can only be implemented reliably and with low expense, if the development of pitting can be continuously monitored in an automated manner. One suitable solution is the camera system for BSD that was used in this paper. Applying the formula to the pitting dataset, setting the value of α to 1 yields the following results: In all cases, the termination criterion ($d_s \geq D_w * 0.3 * \alpha$) takes effect well before the critical temperature of 70 °C is reached on any BSD (indicating its mechanical failure) but after the occurrence of pitting that can be tolerated. This can be seen in Figure 9, where the defined service life is reached well before the components fail mechanically. Therefore, it can be validated that using image data of pittings on BSD spindles, together with the extended formulation for the life time allows the user to accurately indicate wear on the spindle surfaces and replace the components before its mechanical failure. A promising extension of the system is the combination with further sensor variables, such as the motor current of a directly driven BSD. It must be investigated whether a direct assignment of the image data to the motor current data allows to match the defects in the motor current data. The possibility to use only the motor current in operation without the camera system has great advantages with regard to practical use.

## V. Conclusion

The authors presented the results of an empirical analysis of the visually detectable wear progress on ball screw drive (BSD) spindles based on image data. The dataset was generated using an integrated camera system presented in (Schlagenhauf et al., 2019). In contrast to state-of-the-art wear identification approaches that resort to indirect wear effects, the image-based visual analysis, on the other hand, marks the first direct approach. We have shown that the geometric properties of individual pits can be precisely analyzed in image data. Consequently, the wear condition of a BSD can be reliably quantified from images with a high spatial resolution. In addition, it has been shown that pits are reliably detected at a very early stage, allowing the entire wear progression from the initial formation of new pits to the growth of existing pits to be captured and analyzed. Hereby, the measurement uncertainty remains almost constant over the entire service life, whereby only the deposits of wear particle-enriched lubricant can influence the evaluations of the image data in the very last stages of wear.

The approach presented here is particularly relevant for those applications where wear on the spindle is assumed to be the critical wear factor. For this reason, the approach presented here is not intended to totally substitute existing approaches. Nor is the approach presented here intended to be generally more suitable than existing approaches. Though, especially for the detection of pittings on the spindle of the BSD does the approach described here offer advantages over the state of the art.

In further experiments, the image data could be combined with other non-visual signals to increase their interpretability and further reduce the overall uncertainty of wear quantification.

Schlagenhauf *et al.*: Analysis of the Visually Detectable Wear Progress on Ball ScrewsStamboliska, Z., Rusiński, E., & Moczko, P. (2015). *Proactive Condition Monitoring of Low-Speed Machines*. *SpringerLink Bücher*. Springer International Publishing; Springer. http://search.ebscohost.com/login.aspx?direct=true&scope=site&db=nlebk&AN=906359 https://doi.org/10.1007/978-3-319-10494-2

Stockinger, M. (2011). *Untersuchung von Methoden zur Zustandsüberwachung von Werkzeugmaschinenachsen mit Kugelgewindetrieb* [Dissertation, Friedrich-Alexander-Universität Erlangen-Nürnberg, Erlangen]. Deutsche Nationalbibliothek.

Touret, T., Changenet, C., Ville, F., Lalmi, M., & Becquerelle, S. (2018). On the use of temperature for online condition monitoring of geared systems – A review. *Mechanical Systems and Signal Processing*, *101*, 197–210. https://doi.org/10.1016/j.ymssp.2017.07.044

Veith, M., Zimmermann, A. W., Hillenbrand, J., & Fleischer, J. (2020). Detektion des Vorspannungsverlusts in Kugelgewindetrieben/Detection of preload loss in ball screw drives@ Optimization of machine tool maintenance with the Guard Plus system. *Wt Werkstattstechnik Online*(07/08), 485–490.

Verl, A., Heisel, U., Walther, M., & Maier, D. (2009). Sensorless automated condition monitoring for the control of the predictive maintenance of machine tools. *CIRP Annals*, *58*(1), 375–378. https://doi.org/10.1016/j.cirp.2009.03.039

Xi, T., Kehne, S., Fujita, T., Epple, A., & Brecher, C [Christian] (2020). Condition Monitoring of Ball-Screw Drives Based on Frequency Shift. *IEEE/ASME Transactions on Mechatronics*, *25*(3), 1211–1219. https://doi.org/10.1109/TMECH.2020.2969846

Yagmur, T. (2014). *Analyse, Verbesserung und Beschreibung des Verschleißverhaltens von Kugelgewindetrieben für Werkzeugmaschinen: Ergebnisse aus der Produktionstechnik* [Dissertation]. RWTH Aachen, Aachen.

Yang, H [Hanbo], Jiang, G., Sun, Z., Zhang, Z., Zhao, F., Tao, T., & Mei, X. (2021). Remaining Useful Life Prediction of Ball Screw Using Precision Indicator. *IEEE Transactions on Instrumentation and Measurement*, *70*, 1–9. https://doi.org/10.1109/TIM.2021.3087803

Yang, Q., Li, X., Wang, Y., Ainapure, A., & Lee, J. (2020). Fault Diagnosis of Ball Screw in Industrial Robots Using Non-Stationary Motor Current Signals. *Procedia Manufacturing*, *48*, 1102–1108. https://doi.org/10.1016/j.promfg.2020.05.151